\documentclass[prb,twocolumn]{revtex4}
\usepackage{graphicx}

\begin{document}

\title{Time-resolved spectra of polar-polarizable chromophores in solution}

\author{Francesca Terenziani}
\email{terenzia@nemo.unipr.it}
\affiliation{Dip. di Chimica GIAF
 Universit\`{a} di Parma, 43100 Parma, INSTM UdR Parma, Italy}
\author{Anna Painelli}
\email{anna.painelli@unipr.it}
\affiliation{Dip. di Chimica GIAF
 Universit\`{a} di Parma, 43100 Parma, INSTM UdR Parma, Italy}

\begin{abstract}
A recently proposed model for steady-state spectra of polar-polarizable 
chromophores is extended to describe time-resolved spectra.
The model, based on a two-state picture for the solute and on a
continuum overdamped description for the (polar) solvent, grasps the
essential physics of solvation dynamics, as demonstrated by the comparison
with experimental spectra. The solute 
(hyper)polarizability is responsible for 
spectroscopic features that cannot be rationalized within the
standard picture based on a linear perturbative treatment 
of the solute-solvent interaction. 
In particular, the temporal evolution of band-shapes
and the appearance of temporary isosbestic points, two common
puzzling features of observed spectra, are natural consequences
of the molecular hyperpolarizability and of the consequent coupling
between solvation and vibrational degrees of freedom.
\end{abstract}

\maketitle

\section{Introduction}

 Time-resolved spectroscopy is a powerful tool to investigate solvent 
dynamics \cite{simon,fleming},
one of the central issues in modern chemical physics, and a key to 
understand the rate of chemical reactions in solution \cite{rossky,heitele}. 
Detailed information on solvent dynamics can in fact be obtained 
from time-resolved spectra of
properly chosen {\it probe molecules} in solution. Good solvation 
probes have strongly solvatochromic and intense absorption (and/or emission)
bands in the visible region \cite{reichardt}, as not to interfere
with  transitions in the ultraviolet region. 
Polyconjugated chromophores with 
good electron-donor (D) and acceptor (A) groups 
(also called push-pull chromophores)
work fine in this respect: intense and  low-lying transitions are guaranteed 
by polyconjugation, and the large charge displacement from D to A 
(or viceversa) driven by the transition is responsible for a large
variation of the dipole moment
upon excitation, that makes the frequency of the transition 
largely dependent on the polarity of the surrounding medium. 

In the simplest model for solvation, the solvent is described as a 
dielectric continuum medium hosting solute molecules into
Onsager-type cavities \cite{liptay}. 
The electric dipole moment of the (polar) solute
polarizes the solvent so that the solute itself experiences an 
electric field, the reaction field, proportional to the 
solute dipole moment via a proportionality constant that depends on 
the macroscopic properties of the solvent, and on the shape and size
of the Onsager cavity. Two contributions to the reaction field can be
distinguished \cite{dibella,chandler}. 
The electronic clouds of solvent molecules
surrounding the solute deformate in response to the solute dipole moment.
The resulting electronic polarization is very fast with 
respect to the relevant degrees of freedom of the solute. As a consequence, 
the dynamics of the electronic polarization is irrelevant to 
our discussion \cite{chemphys}.
The second contribution to the reaction field originates from the reorientation
of the (polar) solvent molecules around the solute. This contribution, 
that of course vanishes in non-polar solvents, is slower than the solute 
degrees of freedom \cite{liptay,dibella}. 
It is responsible for the large solvatochromic 
behavior of push-pull chromophores \cite{reichardt} 
and dominates the temporal evolution of observed spectra in 
time-resolved experiments \cite{simon}.

In a typical time-resolved experiment the system is shot by
a pulse, or a series of pulses, that induces a transition between the
ground and the excited state (or viceversa). 
  During this process the solvent molecules
have not time to reorient  in response to the sudden change of the
solute dipole moment, and an out of equilibrium (orientational) polarization 
is generated. Provided the state generated by
impulsive excitation lives long enough, the orientational 
contribution to the reaction field ($F_{or}$) slowly evolves from its 
initial value (i.e. that equilibrated with the solute dipole moment 
before excitation), towards the equilibrium value 
for the state reached upon excitation. 
Within linear  perturbation theory, the energy of the states 
involved in the transition is  linearly affected by the reaction field,
and the frequency of the relevant transition acquires a linear 
dependence on the reaction field: 
$\hbar\omega(F_{or})=\hbar\omega(F_{or}=0) - F_{or}(\mu_E-\mu_G)$, 
where $\mu_{G(E)}$ is the electric dipole moment 
of the ground (excited) state \cite{liptay}. 
Then, the experimentally accessible temporal evolution 
of the transition frequency, $\omega(t)$,
gives direct information on the temporal evolution of
the orientational solvent polarization.

The continuum dielectric model reduces the highly irregular and 
complex motion of the solvent molecules to the motion of
a single effective solvation coordinate,  the 
reaction field \cite{loring}.
The dynamical behavior of this special coordinate
is subject to large frictional random forces, that guarantee
for an energy dissipation mechanism \cite{heitele}. 
In the hypothesis of large friction (overdamped motion) the resulting
dynamics is very simple.
The solvation correlation function 
(also called the spectral correlation function),
\begin{equation}
C(t)= \frac{\omega(t)-\omega(\infty)}{\omega(0)-\omega(\infty)}
\label{cdit}
\end{equation}
has in this hypothesis a simple exponential form, and the corresponding 
relaxation time, $\tau_S$, coincides with the 
longitudinal relaxation time, $\tau_L$, of the pure solvent \cite{simon}.
Deviations from this simple behavior are expected in the case of
low-friction or when the translational motion of solvent molecules 
dominates over orientational motion \cite{simon,zwan}.

Extensive experimental work has been devoted by several groups to collect
time-resolved spectra of polar chromophores in 
solution \cite{lds750b,maroncelli2,rosenthal,jarzeba,maroncelli1,dcm2,dcm1,dcm,lds750,c153,ernsting,blanchard}.
The interpretation of experimental data proved
more difficult than described in the simple picture summarized above. 
First of all 
a finite contribution to the Stokes shift ($\omega(0)-\omega(\infty)$) 
is given 
by the  vibrational degrees of freedom of the
solute molecule, as demonstrated most clearly by the observation of 
finite Stokes shifts for chromophores dissolved in non-polar solvents. 
Internal vibrational coordinates are in fact  slower than electronic degrees
of freedom, and hence contribute to the Stokes shift. However,
thanks to the different timescales of vibrational and solvation motions,
their contributions to the temporal evolution of  spectra can be 
at least approximately separated \cite{maroncelli1}.
The basic idea is that, after impulsive excitation, 
vibrational degrees of freedom relax almost
instantaneously with respect to solvation time-scales,
so that vibrational relaxation is essentially done
before  solvent relaxation starts.
Accordingly, the zero-time frequency, $\omega(0)$, 
to be inserted into Eq.~(\ref{cdit}), is not the 
true zero-time frequency, i.e. that relevant to the  unrelaxed 
(vertical) system, but represents  the frequency measured at an 
 effective zero-time for the solvation motion, i.e. the
hypothetical time  when vibrational coordinates are 
fully relaxed, but the solvent  is still frozen in its vertical configuration. 
This effective $t=0$ time, first introduced by Maroncelli and
coworkers in the analysis of time-resolved fluorescence spectra \cite{fee},
 has no experimental counterpart, and $\omega(0)$
cannot be measured directly. Maroncelli proposed a clever approach to 
extract $\omega(0)$ from experimental steady-state spectra
collected in polar and apolar solvents \cite{fee}.
The procedure unavoidably introduces uncertainties in the estimated 
$\omega (0)$ that
add to the uncertainties in the  frequencies estimated from the
broad and asymmetric fluorescence bands \cite{maroncelli2}. 
Precise estimates of $C(t)$ are therefore difficult.
On a more fundamental vein, the simplest solvation model outlined above, and,
{\it a fortiori}, the approach suggested by Maroncelli, both rely on
the hypothesis of instantaneous vibrational relaxation and on 
the hypothesis of an ideal solvation probe: i.e. of a molecule
whose spectral properties are largely affected by the surrounding medium,
but whose presence does not affect  the solvent 
dynamics. The assumption of a complete separation of vibrational 
and solvation time-scales is expected to work well unless one is
interested in the very early times of solvation (the first few tens of
femtoseconds). Deviations can be predicted for molecules with very
slow (conformational) degrees of freedom, particularly if solvents
with very fast relaxation dynamics are considered.
The assumption of the solvent response being unaffected by the
perturbing solute is only valid in the linear regime for
the solute-solvent perturbation.
But linear approaches are inadequate to treat solvation of push-pull 
chromophores.
Just in view of the strong absorption in the visible region, in 
fact, push-pull chromophores have large polarizabilities (and 
hyperpolarizabilities): the reaction field then not only affects the solute
energy, but also its dipole moment. The dipole moment in turn affects
the reaction field in a feed-back mechanism that is responsible for 
large non-linearity. 

In recent years we have developed a simple yet powerful method
to describe steady-state electronic and vibrational spectra of polar
chromophores in solution \cite{cpl,sm1,jpc1,baba}. 
The solute is described in terms of two electronic
states linearly coupled to internal vibrations (Holstein coupling) 
and to the reaction field \cite{chemphys}. 
This model allows for non-perturbative solutions and
naturally describes several features of experimental 
spectra that were not understood in standard linear approaches. 
The reliability of the model is proved by extensive comparison with 
steady-state electronic and vibrational spectra
collected for several chromophores in different 
solvents \cite{baba,jpc2,sm2}. 
In this paper we extend the model to describe time-resolved spectra.
In particular we show that accurate time-resolved spectra
of polar-polarizable chromophores can be calculated adopting the
same set of parameters used for steady-state spectra
without the need to introduce additional adjustable parameters. 
Working in the hypothesis of a complete separation of solvation and
vibrational dynamics, the model
accounts for deviations of $C(t)$ from the simple exponential form,
and rationalizes the observation of time-dependent band-shapes
and the appearance of temporary isosbestic points (TIPs).
These characteristic features of time-resolved spectra of polar-polarizable
chromophores \cite{dcm,lds750,c153,ernsting,blanchard}, 
are a direct consequence of the solute (hyper)polarizability, and
cannot be rationalized within the standard linear picture 
for solvation outlined above. 
As a matter of fact these features have often been taken as
evidences of the failure of the continuum solvation
model \cite{maroncelli2,rosenthal,jarzeba} or
for the failure of the two-state description of the 
chromophore \cite{blanchard}. 
We demonstrate instead that they can appear as natural consequences 
of the molecular polarizability and can be described 
within the simplest model for solvation, 
provided   the fundamental interactions 
of electronic  degrees of freedom with  molecular vibrations and 
with the solvent reaction field are treated beyond linear perturbative
approaches.

\section{The model}

 We consider two electronic states linearly coupled to molecular 
vibrations and to solvent degrees of freedom, according to the following 
Hamiltonian \cite{chemphys}:
\begin{eqnarray}
{\mathcal{H}} &=& 2z_0\hat{\rho} - t \sum_{\sigma}\left(a^{\dag}_{D,\sigma}
a_{A,\sigma}^{}+a^{\dag}_{A,\sigma}a_{D,\sigma}^{}\right) \nonumber \\
&+& \sum_i \left[\frac{1}{2}(P_i^2+ \omega_i^2Q_i^2)-
\sqrt{2\epsilon_{i}}\omega_i Q_i\hat{\rho}\right]
\label{hamiltonian}
\end{eqnarray}
where the first two terms describe the electronic problem, with 
$\hat{\rho}=\sum_\sigma a_{A,\sigma}^\dagger  a_{A,\sigma}$
 counting electrons on A site, and $a_{A,\sigma}^\dagger$,
$a_{D,\sigma}^\dagger$ creating an electron with spin $\sigma$ on D and 
A site, respectively.
$Q_i$ and $P_i$, with $i >0$, are the coordinates 
and conjugated momenta describing the  
coupled vibrations of the chromophore, 
and $Q_0$ is the effective solvation coordinate.
All coordinates are  harmonic, with frequency 
$\omega_i$; the relevant relaxation energy is $\epsilon_i$.
In the following we explicitly account for a single coupled vibration,
$Q_1$. This simplifies equations and limits the number of adjustable
parameters in the model, with no major loss of accuracy in the 
calculation of electronic spectra \cite{baba}.
The extension to the multimode case is possible, and interesting effects
of the mixing among different vibrational coordinates are indeed
expected in vibrational spectra \cite{jpc1,sm2}.

In the adiabatic approximation, the Hamiltonian in Eq.~(\ref{hamiltonian}) 
can be diagonalized at fixed $Q_i$'s on the basis of $|DA\rangle$ and
$|D^+A^-\rangle$ states \cite{chemphys,cpl98}.
The resulting ground and excited states are
fully defined in terms of a single $Q_i$-dependent parameter, 
$\rho=\langle G|\hat \rho |G\rangle$, as follows:
\begin{eqnarray}
 |G\rangle &=&\sqrt{1-\rho} |DA\rangle +\sqrt{\rho} |D^+A^-\rangle \nonumber \\
 |E\rangle &=&-\sqrt{\rho} |DA\rangle + \sqrt{1-\rho}|D^+A^-\rangle
\label{twostate}
\end{eqnarray}
The corresponding energies define the  multidimensional 
ground and excited state potential energy surfaces (PES), as shown 
in Fig.~1a. 
The calculated
PES are interesting in several respects. All coordinates inserted into
the Hamiltonian  are harmonic, and no term appears into Eq.~(\ref{hamiltonian})
explicitly accounting for the mixing of different coordinates. 
The PES in Fig.~1 are instead clearly 
anharmonic and indicate a large mixing between the two involved coordinates.
The softening of the ground state PES, and the hardening of the excited state
PES are also recognized.
All these effects are due to the coupling of the $Q_i$'s to the 
electronic system \cite{jpc1}. In particular we underline that the
coupling between vibrational and solvation
coordinates stems out naturally from their
common coupling to the electronic system.

The PES in Fig.~1 represent the potential for the motion of slow coordinates,
$Q_i$'s.  Internal vibrations with typical frequencies in the mid-infrared
region, are true quantum-mechanical objects. 
The relevant anharmonic problem defined by the PES in Fig.~1 has been 
solved numerically, and large effects of the anharmonicity of internal 
vibrational modes have been demonstrated in static 
hyperpolarizabilities \cite{cplfreo,jcp}. On the contrary, vertical
processes, like, e.g.,  absorption and emission are hardly affected
by anharmonicity \cite{cplfreo,tpa,under}. 
In particular, vertical processes  are quite accurately 
described by replacing the two anharmonic PES by the two parabolas that best 
fit the ground and excited state PES in the relevant region of the 
$Q$ space (best harmonic approximation \cite{cplfreo,jcp,tpa}).
Of course the role of anharmonicity can be  recognized in the variation of the 
effective harmonic parameters fitting absorption and emission processes in this
local-harmonic approximation.

Within the local harmonic 
approximation, electronic spectra and their variation in the $Q$-space 
are fully defined in terms of $\rho$. 
The vertical transition frequency and the transition dipole moment 
are \cite{cpl,jpc1}: 
\begin{eqnarray}
\hbar \omega_{CT}&=&\frac{\sqrt{2} t}{\sqrt{\rho(1-\rho)}} \nonumber\\
\mu_{CT}&=& \mu_0\sqrt{\rho (1-\rho)} 
\label{spectra}
\end{eqnarray}
where $\mu_0$ is the dipole moment of $|D^+A^-\rangle$.
The Huang-Rhys (HR) factor for the coupled vibrational coordinate
($Q_1$) is \cite{cpl,jpc1}:
\begin{equation}
\lambda _1 = \frac{\epsilon_1}{\hbar \omega_1} (1-2\rho)^2
\label{hr}
\end{equation}
Based on  Eqs.~(\ref{spectra}) and (\ref{hr}), 
absorption and emission spectra at any specified 
position in the $Q$ space are calculated as \cite{baba}:
\begin{equation}
{\mathcal{S}}(\omega)\propto \omega^M 
|\mu_{CT}|^2\sum_n \frac{\lambda_1^2}
{\sqrt{n!}}I(\omega-\omega_n)
\label{shape}
\end{equation}
where $\omega_n$ is the frequency of the $n$-th vibronic line, and $I(\omega)$ 
is the intrinsic line-shape function that we choose as a Gaussian with 
standard deviation $\sigma$.
$M=1,3$ for absorption and emission, 
respectively.

The solvation coordinate describes the slow orientational motion
of polar solvent molecules around the polar solute. The relevant coordinate,
$Q_0$, can then be treated as a classical coordinate, and, at  finite 
temperature, it is responsible for inhomogeneous broadening. 
Since the solvation motion is coupled to electronic degrees of
freedom, and, through them, to internal vibrations, inhomogeneous broadening 
has in these systems quite peculiar features. 
Due to thermal disorder in fact, the solution
can be thought of as a collection of chromophores each one surrounded by a
slightly different $Q_0$ configuration. Both electronic and vibrational 
motions are very fast if compared to solvation, so that each solute molecule 
readjusts instantaneously  its polarity and its geometry 
in response to  the local $Q_0$ configuration \cite{jpc1}. 
Thermal disorder in $Q_0$ 
is then responsible for disorder in electronic and vibrational 
properties, that shows up in electronic and vibrational spectra, and,
most apparently, in the anomalous dispersion of resonant Raman frequencies 
with the excitation line \cite{jpc1,jpc2}.
With reference to Fig.~1, the thermally equilibrated state 
in either the ground or excited state PES is defined in terms of a Boltzmann 
distribution of states lying along the dashed lines drawn across the
isopotential lines in Fig.~1b and 1c to mark the local minimum path
with respect to the vibrational coordinate.  The steady-state absorption
and emission spectra are then sums of the contributions in
Eq.~(\ref{shape}) calculated for each point along the minimum path
in the ground state and excited state PES, respectively, 
and weighted by the relevant Boltzmann
distribution \cite{jpc1,baba}.

This simple model has been applied successfully to describe 
electronic and vibrational spectra of several chromophores, reproducing many 
features of experimental spectra that were not understood 
before \cite{baba,jpc2,sm2}.
Just as an example, Fig.~2A and 2B shows the absorption and emission spectra
calculated in the proposed approach to fit experimental spectra of 
DCM dissolved in several solvents. 
These spectra accurately reproduce 
experimental data \cite{baba}, and deserve some comment. 
The red-shift of absorption and emission bands
with increasing solvent polarity (positive or normal solvatochromism) 
is typical of chromophores
with a neutral ground state ($\rho < 0.5$) \cite{reichardt}.
For these chromophores the solute
polarity increases upon excitation, and the excited state is stabilized
by a polar surrounding medium. 
The evolution of the band-shape with the solvent polarity is a more subtle
phenomenon that can be understood only if the polarizability of the solute is
accounted for \cite{jpc1,baba}. 
In fact, the polarity of the (polarizable) solute increases
 with the polarity of the surrounding medium. For neutral chromophores then
$\rho$ increases towards 0.5, and HR factors decrease: even if 
inhomogeneous broadening increases with the solvent polarity,
absorption and emission bands apparently narrows, since 
the underlying vibronic profile narrows ($\lambda_1$ decreases, see
Eq.~(\ref{hr})). For zwitterionic chromophores 
($\rho >0.5$), instead, the ionicity increases towards 1, and HR factors
 increase with the solvent polarity: 
in this case the effects of solvent polarity on inhomogeneous broadening
and HR factors sum up to widen absorption and emission bands.
This behavior is experimentally verified 
in dyes with inverse solvatochromism \cite{lds750,metzger,hr},
and is reproduced in Fig.~2C and 2D for a hypothetical dye with the same 
molecular parameters as DCM, but a different $z_0$, 
as to have a zwitterionic ground state.

Another direct consequence of the molecular polarizability
is recognized in the non-specularity of
steady-state absorption and emission bands \cite{cpl,baba}.
The behavior is similar 
for neutral and zwitterionic chromophores, with narrower emission than 
absorption bands. In all chromophores, in fact, the mixing between the two 
basis states increases in the relaxed excited state, so that, in both 
cases  steady-state emission spectra correspond to states with $\rho$ nearer 
to 0.5 than states relevant to absorption spectra. 
According to Eq.~(\ref{hr}),
HR factors for emission are always smaller than for absorption then justifying 
quite naturally the observed narrowing of emission bands.

\section{Relaxation dynamics}

The success of the two-state model in reproducing steady-state absorption and 
emission spectra invites modeling  time-resolved experiments,
as related to the dynamics of non-equilibrium states created by 
impulsive excitation in either the ground or excited state PES. 
We are not interested in the fast relaxation of vibrational coordinates 
(typically accomplished within few tens of femtoseconds after 
excitation) \cite{maroncelli1,ernsting}, 
rather we model the slower solvation dynamics in the hundreds 
of femtoseconds to picosecond temporal window. 
The problem is therefore greatly simplified if we work in the
hypothesis of a 
complete separation of vibrational and solvation dynamics.
First of all, just after
a (sequence of) pulse(s) has created an out of equilibrium state in one of
the PES, an instantaneous relaxation of  vibrational 
coordinates takes place before solvation relaxation starts \cite{maroncelli1}. 
Moreover, during the subsequent 
relaxation of solvation degrees of freedom,
internal coordinates immediately readjust with respect
to the instantaneous configuration of the solvent. 
Due to the non-linearity of the 
coupled problem, the electronic wavefunction readjusts
 following $Q_0$ and, as a consequence, the equilibrium geometry of 
the chromophore varies
with $Q_0$. Since electronic and vibrational degrees of freedom are 
much faster than solvation, they are always in equilibrium with the
instantaneous configuration of the solvent, and the relaxation
of the effective solvation coordinate does not proceeds along a horizontal 
line (fixed $Q_1$) in Fig.~1b or 1c, but follows the dashed lines 
drawn across the isopotential lines to mark 
the minimum energy path in each surface.

In a time-resolved fluorescence experiment, 
an impulsive excitation drives the system from the thermally equilibrated 
ground state (centered on point G in Fig.~1b) to the vertical 
excited state (point G' in Fig.~1c). 
After excitation, the instantaneous relaxation of vibrational 
degrees of freedom drives the system 
to point F, i.e. to the local minimum in the excited state PES, 
at  the $Q_0$ relevant to the equilibrium ground state configuration. 
During the G$\rightarrow$G'$\rightarrow$F path, the $Q_0$ configuration
is unaffected.
Once the system has reached point F, solvent relaxation starts.
Since vibrational degrees of freedom follow adiabatically 
the solvent relaxation, the relaxation path towards the minimum of the 
excited state PES (F$\rightarrow$E) follows the minimum energy path, i.e.
the dashed line drawn across points F and E in Fig.~1c. So, in spite of
 being driven by orientational degrees of freedom of the solvent,
the dynamics of the system occurs along an effective coordinate that
describes a concerted motion of solvation degrees of freedom
and internal vibrations. 
Once again, the coupling between the two kinds of motion originates from 
their common coupling to the electronic system, and is a direct consequence
of the molecular polarizability. 

Time-resolved fluorescence spectra can be calculated following the same 
procedure described in the previous section for steady-state spectra, provided
the temporal evolution of the $Q_0$ probability distribution is known.
To such an aim we rely on a Fokker-Planck description of the dynamics of the
joint probability distribution in the space of the solvation coordinate and
momentum \cite{caldeira,mukamel}, $W(Q_0,P_0)$,
as described by  the following equation of motion:
\begin{equation}
	\frac{\partial W}{\partial t} = 
	-\dot{Q_0} \frac{\partial W}{\partial Q_0}
	-\dot{P_0}\frac{\partial W}{\partial P_0}+
	\gamma\left\{W+P_0\frac{\partial W}{\partial P_0}+
	kT\frac{\partial^2 W}{\partial P_0^2}\right\}
\label{fokker}
\end{equation}
where $\gamma$ is the phenomenological friction coefficient for
the motion along the relevant relaxation path. 
Starting from point F, where the distribution is rigidly 
translated from the equilibrium ground state distribution, 
the Fokker-Planck equation can be easily integrated along 
the F-E path in Fig.~1c. 
The calculation can be simplified by recognizing the
overdamped nature of the solvation coordinate ($\gamma \gg \omega_0$). 
In the overdamped regime,  the momentum rapidly attains
the equilibrium, so that it needs not to be considered as an independent
dynamic variable \cite{mukamel}. 
The only relevant dynamics involves the $Q_0$-distribution,
$w(Q_0)$, as described by Smoluchowski equation:
\begin{equation}
	\frac{\partial w}{\partial t} = \frac{1}{\gamma} 
	\left[ w\frac{\partial^2 V}{\partial Q_0^2} + 
	\frac{\partial V}{\partial Q_0} \frac{\partial w}{\partial Q_0} + 
	kT\frac{\partial^2 w}{\partial Q_0^2} \right]
\label{smolu}
\end{equation}
where $V(Q_0)$ is the (classical) potential for the $Q_0$ motion.
We verified that for $\gamma >2 \omega_0$ the Smoluchowski equation
leads to the same results as the Fokker-Planck equation. 
Apart from  a simplification in the integration procedure,  working in the 
overdamped limit has  the advantage of reducing the 
number of free parameters required to model solvation dynamics. 
In fact, as it is well known \cite{mukamel}
and as we have explicitly verified, in that limit all dynamic properties
depend on $\gamma$ and $\omega_0$ only through the ratio 
$\tau=\gamma/\omega_0^2$. $\tau$  represents an effective solvation time, 
and can be fixed as the longitudinal relaxation time ($\tau_L$)
of the pure solvent \cite{simon}.

Fig.~3 shows time-resolved spectra calculated for DCM
 dissolved in two different solvents (CH$_3$CN and CHCl$_3$). 
All  molecular parameters, and the solvent relaxation energies as well, are
kept fixed at the values previously obtained from the fit of steady-state 
spectra \cite{baba} (cf Fig.~2, left panels). 
According to literature data, we fix
$\tau= 0.3$ and 2.8 ps for CH$_3$CN and CHCl$_3$, 
respectively \cite{maroncelli1}. 
The spectra  in Fig.~3 compare nicely with
 experimental spectra \cite{dcm2,dcm1}, and accurately reproduce 
 the progressive red-shift  of the band, as well as its narrowing.
In order to better understand the origin of the evolution of
band-shapes during solvation, Fig.~4 shows time-resolved spectra 
calculated for the same parameters as Fig.~3A, but without accounting for
inhomogeneous broadening, i.e. by collapsing the $w(Q_0)$ 
distribution into a $\delta$-function.
The comparison between inhomogeneously broadened spectra in Fig.~3A and  
homogeneous spectra in Fig.~4 demonstrates that  the evolution of
the band-shape accompanying the solvent relaxation 
is essentially due to the evolution of the vibronic structure
(i.e. of HR factors), and reflects quite naturally the 
evolution of the molecular geometry along the relaxation path.
The $w(Q_0)$ distribution, obtained from the solution of the Smoluchowski 
equation, also evolves with time, but it affects calculated spectra  
 in a more subtle way. In fact $w(Q_0)$ is responsible for
inhomogeneous spectral broadening, whose temporal evolution is hardly
recognized in the fairly broad emission spectra of polar chromophores in 
polar solvents.

Similar features, including the progressive red-shift of the 
emission band and its narrowing, are recognized in time-resolved 
emission spectra of many other dyes, including neutral dyes like
coumarin 153 \cite{maroncelli1}, and zwitterionic dyes 
like LDS-750 \cite{lds750b}.
In fact, in spite of the opposite solvatochromism,
neutral and zwitterionic dyes are predicted to have similar behavior with 
respect to time-resolved emission, as demonstrated by the time-resolved 
spectra calculated  in Fig.~5 for the same  zwitterionic dye whose 
steady-state absorption and emission spectra are reported 
in Fig.~2, right panels. 
Both the red-shift and the narrowing of time-resolved
emission bands are due to the increase of the  mixing between the two 
basis states during the relaxation of the system along the effective 
solvation coordinate, as discussed in the previous section with reference 
to steady-state spectra.

Other time-resolved experiments  
can be described within the proposed approach.
Transient hole-burning experiments have been designed to follow
the relaxation along the ground state PES.
Specifically, we make reference to the experiment described 
in Ref.~\onlinecite{ernsting}.
In this experiment, a solution of coumarin 102 (C102) dissolved in CH$_3$CN 
is impulsively excited to transfer population from the minimum 
of the ground state PES (point G in Fig.~1b) 
to the vertical excited state (point G' in Fig.~1c). 
After a delay time long enough as to 
allow for the  complete relaxation of  the system in the excited PES (from 
G' to E in Fig.~1c), a second dump pulse is applied to vertically transfer 
the equilibrated population in the excited state PES into the ground state
PES (from E to E' in Fig.~1b). 
We are not interested in modeling the initial 
relaxation of vibrational coordinates occurring during 
the first few tens of fs \cite{ernsting}, 
but, much as it was done for time-resolved 
fluorescence, we want to model the slower spectral evolution due to the
solvent relaxation. 
We then assume an instantaneous (within the relevant solvation timescale) 
relaxation along the  vibrational coordinate, towards  point R in the 
ground state PES (Fig.~1b) lying on the local minimum path with respect 
to $Q_1$. 
The calculation of transient absorption spectra then goes via the solution 
of the Smoluchowski equation for $w(Q_0)$ from time zero 
(where the equilibrium $w(Q_0)$ relevant to the excited state PES is 
rigidly translated to point R in the ground state PES) 
towards the ground state equilibrium in G.

Figure~6 shows transient differential absorption spectra calculated 
for C102 in CH$_3$CN.  Molecular parameters for C102, as well
as the solvent relaxation energy relevant to C102 in CH$_3$CN, 
are obtained from the fit of steady-state absorption 
and emission spectra \cite{prepa}.
The additional parameter entering the Smoluchowski equation, $\tau$,
is again fixed to 0.3~ps.
In order to compare directly with experimental results, 
 the upper and lower panel in Fig.~6 report the spectra calculated in 
 two different temporal windows. 
The comparison with  experimental spectra in Fig.~5 of 
Ref.~\onlinecite{ernsting} is very good:
in particular we reproduce not only the blue-shift of the band,
but also the appearance of a TIP, and its temporal evolution. 
The origin of TIPs is easily recognized in the
variation of the band-shape of the transient absorption that accompanies
its progressive blue-shift. Whereas the rigid translation to the blue of the 
transient absorption band is quite easily understood based on the simplest
linear model for solvation, the temporal evolution 
of the band-shape is a direct consequence 
of the molecular polarizability. 
In fact, as long as the system relaxes along the R-G path in Fig.~1b, 
the molecule readjusts its polarity (from $\rho\sim 0.23$
to 0.16, in the specific case of C102 in CH$_3$CN), and HR factors 
relevant to the absorption process change with time. 
TIPs are puzzling features of time-resolved spectra that cannot be explained
easily within the standard picture for solvation, but are a natural consequence
of the polarizability of the solute molecule.

TIPs appear quite commonly in pump-probe 
experiments \cite{dcm,lds750,c153,blanchard}. 
In these experiments,
after the application of a pump pulse that drives the system to the 
excited state PES,  the variation of the optical density
of the solution is monitored by a probe-beam as a function of the delay time
after the pulse. The experiment is easier than the hole-burning experiment,
but the interpretation of resulting spectra is more complex. In 
fact at least three processes can contribute to the observed signal at 
each frequency. Both the bleaching of the absorption from the ground state 
(due to the reduced ground state population) and stimulated emission 
decrease the optical density, leading to a negative signal, 
whereas the absorption from the excited state increase the optical density, 
originating a positive signal.
A reasonable separation of the different contributions is only possible
if the three processes occur in well separated spectral regions. 
For polar chromophores in polar solvents, the  large Stokes-shifts 
guarantee in general a reasonable discrimination between signals originating
from the bleaching of the ground state absorption and from 
stimulated emission. 
If the positive signal from excited state absorption does not 
interfere too much, the pump-probe experiment gives information on
the evolution of the emission band during the relaxation of the system along 
the F-E path in the excited state PES (Fig.~1c), so that,
in favorable cases, the pump-probe experiment gives similar information 
as time-resolved fluorescence. According to the previous discussion, we
then predict a progressive red-shift and narrowing of the emission band, 
in quite good agreement with available experimental data for 
several chromophores, with either a neutral \cite{dcm,c153,blanchard} or a
zwitterionic ground state \cite{lds750}. 
We underline once more that the red-shift
of the stimulated emission band is easily rationalized in the standard 
solvation model. On the opposite, the narrowing of the band,
being related to the variation of the molecular polarity with the 
solvent relaxation,  can only be 
understood in models where the molecular polarizability is 
properly accounted for. The concomitant red-shift and narrowing of the 
emission band are responsible for the appearance of TIPs in observed spectra.

\section{Discussion}

Push-pull chromophores are interesting molecules in several respects:
they find application as solvation probes \cite{fleming,reichardt}, 
as laser-dyes, and as chromophores for non-linear optics \cite{kanis}. 
They are also good two-photon absorbers \cite{albota} and
work as molecular rectifiers \cite{metzger2}. 
From a theoretical point of view,
the possibility to describe the electronic structure of these molecules
in terms of an effective two-state model makes these systems fascinating
model systems to understand electron-transfer in solution and specifically
to investigate the physics of solvation in electron-transfer
processes \cite{heitele,barbara,walker,myers}.
In a series of recent papers \cite{chemphys,cpl,jpc1,baba}, we have extended
the standard two-state model for push-pull 
chromophores \cite{oudar,goddard,barzoukas}
to account for solvation and molecular vibration.
The model allows for a non-perturbative treatment of the relevant couplings
and fully accounts for the solute (hyper)polarizability,
yielding to a successful picture for steady-state spectra of
push-pull chromophores in solution \cite{baba,jpc2,sm2}.
Here we extend the model
to time-resolved spectra and demonstrate the important role of
the solute hyperpolarizability in several characteristic and non-trivial
spectroscopic features.

Consider a simple system with no electron-vibration coupling.
Due to the molecular (hyper)polarizability, the reaction field 
affects the solute properties and, in turn, is affected
by the presence of the solute.
As a consequence of this feedback interaction, 
the ground and excited state PES for the $Q_0$ motion
are deformed \cite{under}. 
The linear polarizability ($\alpha_{G/E}$) is responsible
for a renormalization of the relevant harmonic frequencies \cite{jpc1}:
\begin{equation}
\Omega_{G/E} =\omega_0 \sqrt{1-2\epsilon_{0} \alpha_{G/E}/\mu_0^2}
\label{freq}
\end{equation}
while higher order polarizabilities originate anharmonic 
corrections \cite{cplfreo,jcp} that can be fairly large 
for chromophores with large NLO responses.
The role of the molecular first polarizability, $\alpha$, on solvation
dynamics is easily understood and has already been discussed by several
authors \cite{cichos,kim,glasbeek,matyushov}.
If the hyperpolarizability is neglected, the PES are harmonic and 
the solvation relaxation function, $C(t)$, has a simple exponential
form. The relaxation time, $\tau_{G/E}= \gamma/\Omega_{G/E}^2$, 
is however renormalized with respect to the 
relaxation time of the pure solvent, $\tau= \gamma/\omega_{0}^2$,
and depends on specific properties of the solute.
A qualitatively different behavior is expected if 
hyperpolarizabilities are accounted for, leading to anharmonic PES. 
In this case, a non-exponential $C(t)$ is expected, 
whose detailed behavior depends
the details of the relevant PES. The role of the solute 
hyperpolarizabilities on solvation dynamics has not been recognized so far
and the observed deviations of $C(t)$ from the simple exponential behavior
have always been ascribed to the failure of the continuum model for the
solvent or to the underdamped nature 
of the solvation coordinate \cite{maroncelli2,rosenthal,jarzeba}.
Instead a non-exponential behavior can
show up within the simplest model for solvation, 
as a consequence of the solute hyperpolarizability.

The molecular (hyper)polarizability has even more important effects
in the presence of coupled internal vibrations. In fact 
the common coupling of vibrational and solvation degrees of freedom 
to the electronic system induces a mixing between the two kinds of 
coordinates, even if no direct coupling between the modes is included in 
the hamiltonian. The physical origin of the mixing lies in the
molecular (hyper)polarizability: the polarizable molecule in fact 
readjusts its polarity to any variation of solvation or of vibrational
coordinates, but this variation of polarity affects in turn all coordinates,
leading to an interdependence of all motions. As discussed 
in the previous section, even in the simplifying hypothesis 
of a complete separation of 
the timescales of vibrational and solvation motion,
the relaxation of the solvent coordinate always implies, for a polarizable
solute, a concomitant relaxation along the vibrational coordinates. Therefore
time-resolved experiments performed on polarizable chromophores
measure a complex motion where the molecular geometry readjusts 
following the solvent relaxation. Deviations of the observed relaxation time
from the relaxation time of the pure solvent, as well as
deviations from the simple exponential
behavior for $C(t)$, are obviously predicted in this model. 
Just as an example, Fig.~7A compares the  spectral correlation function 
calculated for hole-burning experiment on C102 described 
in the previous section.
Deviations of the exact result from the simple exponential behavior,
as well as an overall slowing down of the solvation dynamics with respect
to the pure solvent, can be appreciated from the figure, even if,
due to the large experimental uncertainties in $C(t)$,
they are possibly too small to be safely identified in experimental data.
Larger effects are predicted for more polarizable solute molecules,
as shown in Figure~7B for phenol blue, an interesting solvatochromic
probe whose large hyperpolarizability is known since date \cite{marder}.
Quite irrespective of their magnitude, these effects are important 
since they demonstrate that the presence of a polarizable solute
affects solvation dynamics: the relaxation time obtained from the
spectral relaxation function is not a property of the pure solvent,
but also depends on the solute molecule.
Good solvation probes should be largely solvatochromic and 
\emph{hardly polarizable} as well.

The two-state model unavoidably assigns the excited state a linear
polarizability equal in magnitude and opposite in sign to the
ground state polarizability, then pointing to a solvation time in the excited
state smaller than the longitudinal relaxation time of the pure solvent.
As discussed in Ref.~\onlinecite{matyushov2} 
the negative polarizability of the excited state
is probably an artifact of the two-state model, that neglects 
the mixing of the low-lying excited state with  higher excited states  
whose (positive) contribution to $\alpha_E$ usually
over-compensate the negative contribution from the mixing 
with the ground state. Estimates of excited state polarizabilities
are available for just a few chromophores and all confirm positive
$\alpha_E$ values \cite{peteneau,matyushov3}.
However, the possibility of observing a negative 
$\alpha$ for the excited state of highly polarizable 
chromophores cannot be excluded a priori. 
In any case, it is possible to modify the
proposed approach to account for the role of (a few) higher excited states. 
However, this would introduce new adjustable parameters
in the model, whose definition is arbitrary 
in the absence of accurate experimental information on the 
polarizability of the  excited state, or on the relaxation dynamics
in the excited state.

The analysis of $C(t)$ is hindered by large experimental uncertainties,
and the effects of the solute (hyper)polarizability are more easily
recognized in the temporal evolution of band-shapes. 
As recently suggested by Matyushov \cite{matyushov}, 
if one accounts for the solute linear polarizability, 
the different curvatures in the ground and excited state PES
lead to different equilibrium $w(Q_0)$ distributions in the two states.
Then a temporal evolution of $w(Q_0)$ is easily predicted
in time-resolved experiments, resulting in an evolution
of the inhomogeneous broadening profiles in observed spectra.
The solute hyperpolarizability makes the phenomenon more
complex: anharmonic PES, as originated by non-linear 
polarizabilities \cite{cplfreo,jcp}, in fact lead to non-Gaussian
distribution profiles. 
However, it is important to recognize that, 
as discussed in the previous section,
band-shapes are actually governed by vibronic profiles, 
inhomogeneous broadening being only responsible for their smearing out. 
The observed evolution of band-shapes is indeed originated by
the coupling between solvation and vibrational degrees of freedom,
as induced by the molecular (hyper)polarizability.
The role of vibronic profiles in
defining observed band-shapes was recognized in Ref.~\onlinecite{marcus},
but the temporal evolution of HR factors was introduced {\it ad hoc}. 
We demonstrate instead that it is a natural consequence of the solute
(hyper)polarizability.

The evolution of band-shapes that accompanies the red- (blue-) shifts
of time-resolved fluorescence (absorption) bands can be responsible
for a crossing of observed spectra.
The resulting TIPs are not true isosbestic points, as their location
smoothly varies with time.
In any case, the appearance of TIPs has puzzled experimentalists and
theorists since date \cite{dcm,lds750,ernsting,blanchard}. 
Recently, a paper appeared aimed to demonstrate
the failure of the two-state picture for push-pull chromophores, 
based on the appearance of TIPs in pump-probe
spectra \cite{blanchard}. 
In this paper, TIPs are ascribed to a barrier-activated 
reaction occurring in the excited state towards a transient emissive 
product. 
Whereas further theoretical and experimental work is needed to settle
this point, we underline that the molecules 
investigated in Ref.~\onlinecite{blanchard} are among 
the best chromophores for 
NLO, and are therefore largely (hyper)polarizable.
The observed large
deviations  of relaxation times from the longitudinal
relaxation times of the pure solvents, point quite nicely to
deviations from the standard linear perturbative picture of the solute-solvent
interaction and perfectly fit within our two-state model for 
polar-polarizable chromophores, and suggest an alternative explanation
of TIPs.

\section{Conclusions}

Time-resolved spectra of polar-polarizable chromophores in solution
are described based on a continuum solvation model.
A two-state picture is adopted for the solute, and linear coupling
to internal vibrations is also accounted for. The model, originally
developed to account for NLO responses \cite{chemphys,cpl98}
and steady-state spectra \cite{cpl,jpc1} of push-pull 
chromophores in solution,
is extended here to describe time-resolved spectra.
The comparison with experimental data confirms
that the proposed model quite successfully grasps the essential 
 physics of polar-polarizable chromophores in solution. 
The model presents some  obvious drawbacks.
The two-state picture must be extended to account for higher-energy states
if detailed information is required on excited state properties.
Moreover, the Holstein model for electron-vibration coupling only accounts for
linear terms: quadratic coupling is expected to be important in cases
when a very different geometry (and hence very different bond orders)
can be attributed to the two basis states. Again the extension of the
model in this direction is easy, but requires detailed spectroscopic
information to fix all the model parameters. Finally, the continuum
model for solvation has serious limitations, particularly in cases were
large site-specific interactions (like, e.g., H-bonds) are expected.
The hypothesis of complete separation of vibrational and solvation
time-scales can fail for molecules with particularly slow internal
degrees of freedom and/or for solvents with a very fast dynamics.
In spite of these limitations, and quite irrespective of its applicability
to some specific systems, the proposed model opens the way to obtain
exact, non-perturbative solutions of the solvation problem also in 
the presence of electron-vibration coupling. The spectral consequences
of the molecular (hyper)polarizability, that is unavoidably
neglected in the standard linear perturbative approach to 
solvation,  are important and non-trivial, and show up quite consistently 
in steady-state and time-resolved spectra of push-pull chromophores.
In particular  here we have 
demonstrated that deviations from the
exponential decay of the spectral relaxation function, 
that are usually taken as proof of the failure of the continuum
model for solvation, appear quite naturally within the simplest picture,
provided the molecular (hyper)polarizability is accounted for. 
Similarly, the temporal
 evolution of band-shapes, as well as the appearance of TIPs, usually 
ascribed to the failure of the two-state model, are another consequence
of non-linearity. More complex models can of course be developed, and will
lead to a better or more detailed description of spectral properties of 
push-pull chromophores in solution. Based on an Occam-razor approach, we
believe however that the 
available body of experimental data does not require 
any more complex physics than described here.

\widetext
\newpage

\begin{figure}
\begin{center}
\end{center}
\includegraphics* [scale=0.8,angle=0]{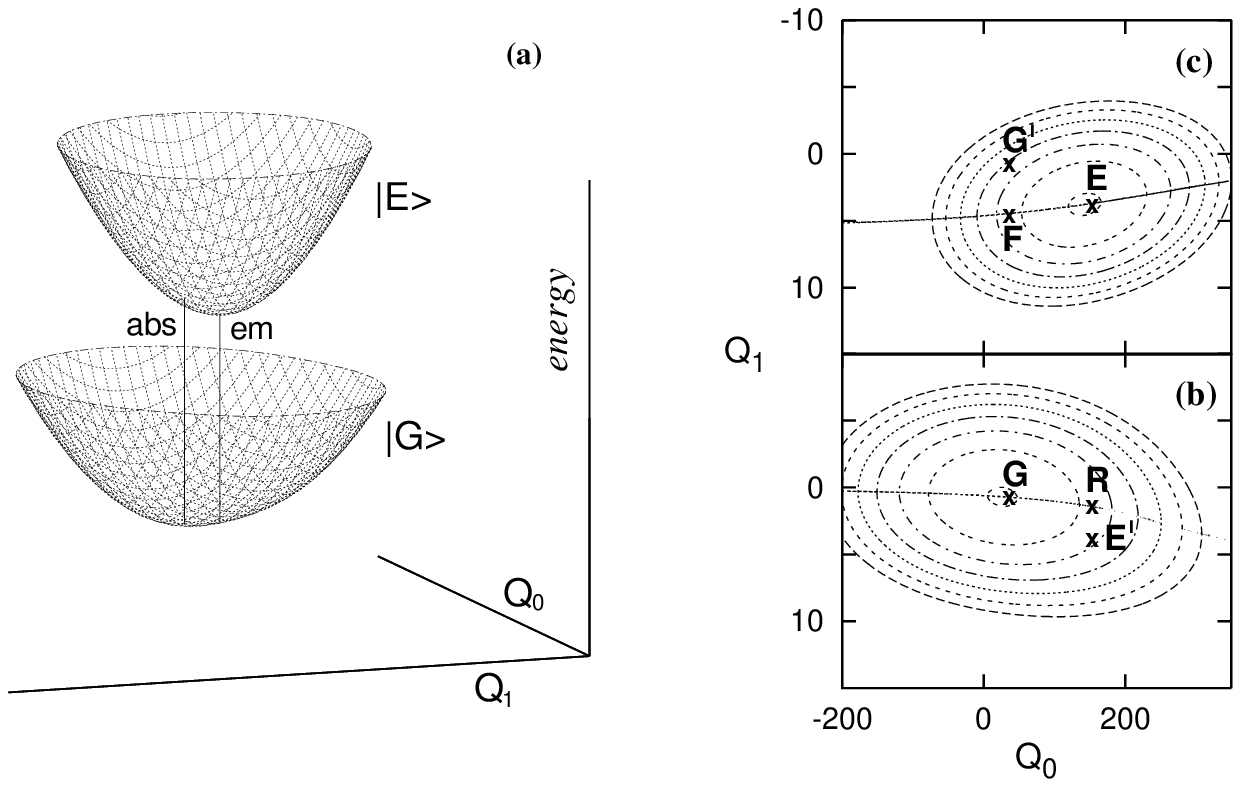}
\caption{(a) Potential energy surfaces calculated for 
the same parameters used below to fit spectra
of DCM in CH$_3$CN: $z_0 = 1.14$~eV, $\sqrt{2}t=0.88$~eV, 
$\epsilon_1=0.45$~eV, $\omega_1=0.16$~eV, $\epsilon_0=0.75$~eV.
$Q_0$ and $Q_1$ describe  the solvation
and vibrational coordinate, respectively. 
The labels {\it abs} and {\it em} mark the 
vertical lines  along which the absorption and emission processes occur.
For the sake of clarity, the energy gap 
between the two surfaces has been enlarged.
(b) Isopotential lines for the ground state and
(c) the excited state PES shown in panel (a).
Lines are relevant to equispaced energy values,
the spacing corresponding to $\omega_1$.
The dotted lines drawn across the isopotential lines mark the 
equilibrium $Q_1$ value as a function of $Q_0$.
Letters on the graphs mark points discussed in the text.
All axes are in arbitrary units.}
\end{figure}

\begin{figure}
\begin{center}
\includegraphics* [scale=0.7,angle=0]{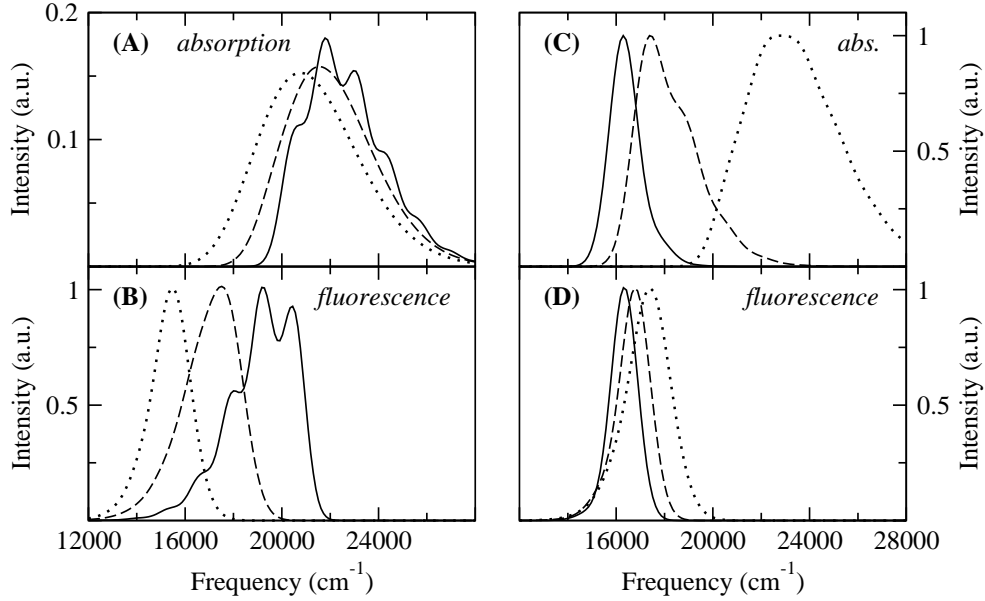}
\end{center}
\caption{Left panels: (A) Absorption and (B) fluorescence spectra calculated 
for DCM (same parameters of Fig.~1) in hexane (continuous lines, 
$\epsilon_0 =0$), CHCl$_3$ (dashed lines, $\epsilon_0 =0.32$~eV),
CH$_3$CN (dotted lines, $\epsilon_0 =0.85$~eV).
Right panels: (C) Absorption and (D) fluorescence spectra calculated 
for a zwitterionic chromophore ($z_0 = 0.1$~eV, $\sqrt{2}t=1$~eV, 
$\epsilon_1=0.5$~eV, $\omega_1=0.18$~eV) for $\epsilon_0=0$ 
(continuous lines), $\epsilon_0 =0.36$~eV (dashed lines)
and $\epsilon_0 =0.85$~eV (dotted lines).
The intrinsic line-width for the Gaussian associated to each
vibronic feature is $\sigma=0.06$~eV.}
\end{figure}

\begin{figure}
\begin{center}
\includegraphics* [scale=0.7,angle=0]{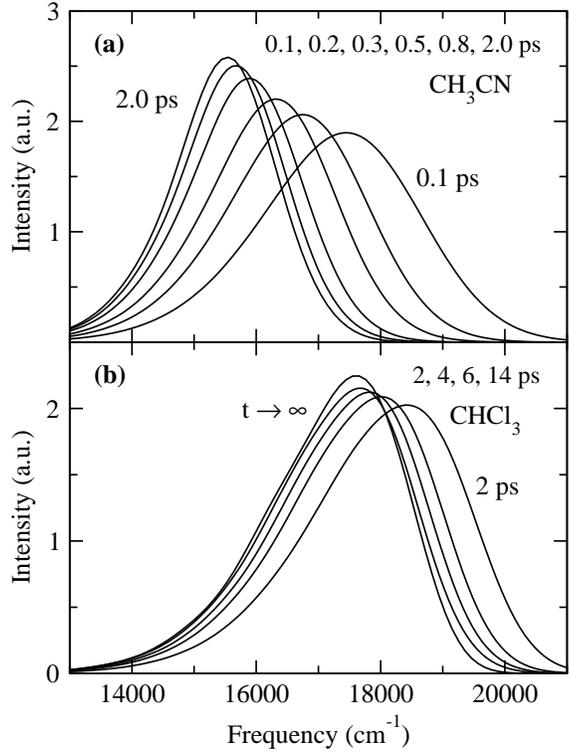}
\end{center}
\caption{Time-resolved fluorescence spectra calculated for
DCM in (A) CH$_3$CN and (B) CHCl$_3$.
The parameters are the same as for Fig.~2, left panels;
$\tau=0.3$ and 2.8~ps in panels A and B, respectively.}
\end{figure}

\begin{figure}
\begin{center}
\includegraphics* [scale=0.8,angle=0]{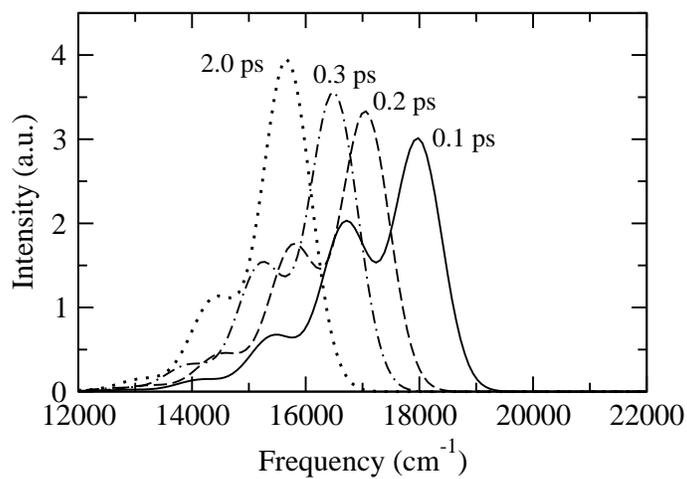}
\end{center}
\caption{The same as Fig.~3A, but without accounting for inhomogeneous
broadening of the spectra.
Again times relevant to the different spectra are marked on the graph.}
\end{figure}

\begin{figure}
\begin{center}
\includegraphics* [scale=0.7,angle=0]{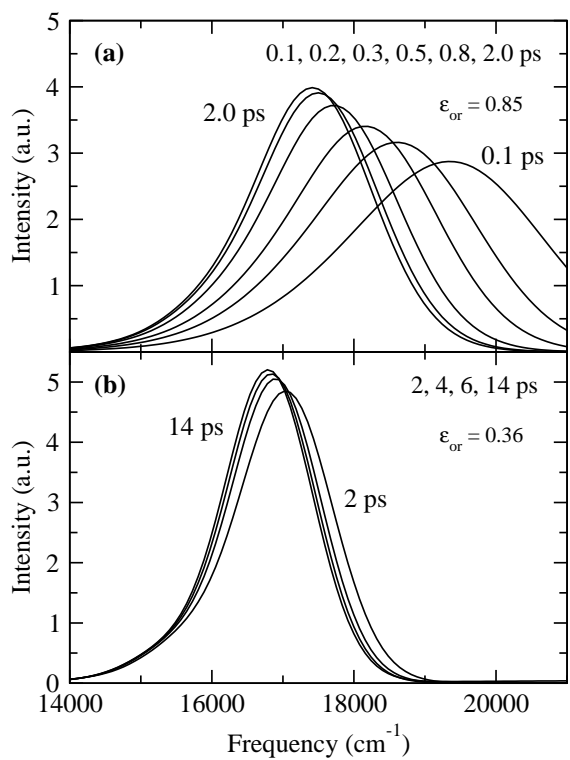}
\end{center}
\caption{The same as Fig.~3, but for the zwitterionic chromophore
of right panels of Fig.~2.}
\end{figure}

\begin{figure}
\begin{center}
\includegraphics* [scale=0.6,angle=0]{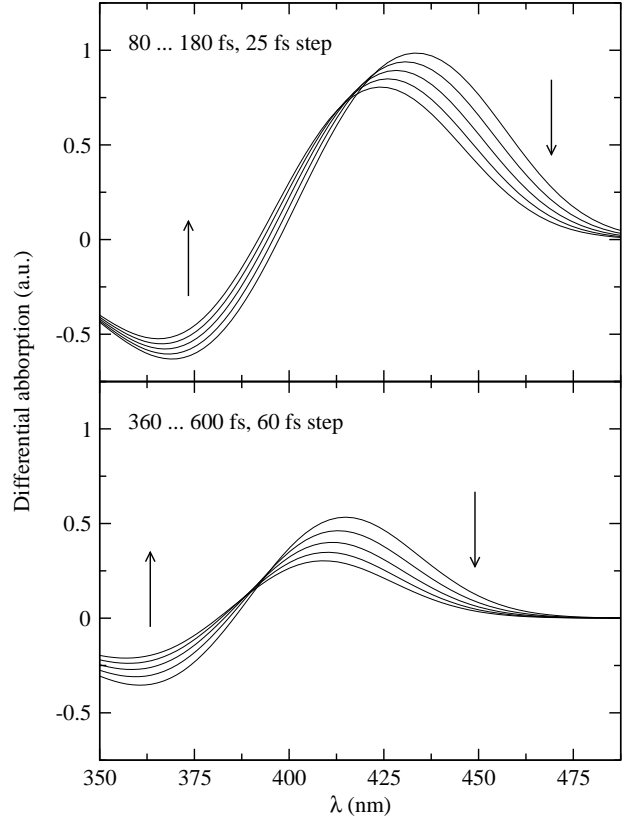}
\end{center}
\caption{Transient differential absorption spectra calculated for 
C102 in CH$_3$CN: $z_0 = 1.28$~eV, $\sqrt{2}t=1.2$~eV, 
$\epsilon_1=0.33$~eV, $\omega_1=0.16$~eV, $\epsilon_0=0.6$~eV.
Spectra are relevant to the times reported on the graphs.
Calculated spectra are shown in two separate windows for a better comparison
with data in Ref.~\onlinecite{ernsting}.
Arrows indicate the direction of increasing time.}
\end{figure}

\begin{figure}
\begin{center}
\includegraphics* [scale=0.6,angle=0]{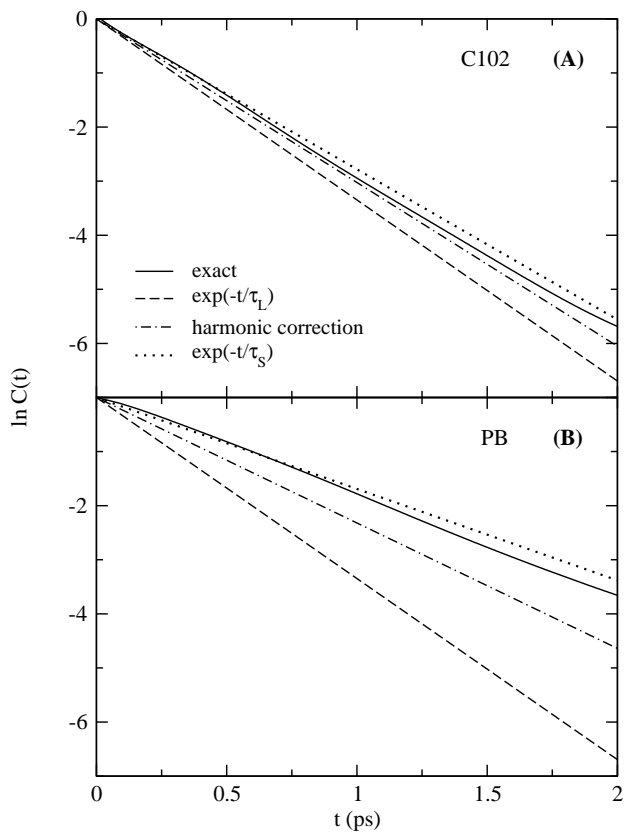}
\end{center}
\caption{Solvation correlation function for the hole-burning experiment,
calculated for 
(A) C102 in CH$_3$CN (the same parameters as in Fig.~6)
and (b) phenol blue in CH$_3$CN ($z_0 = 0.7$~eV, $\sqrt{2}t=1$~eV, 
$\epsilon_1=0.42$~eV, $\omega_1=0.2$~eV, $\epsilon_0=0.7$~eV,
from Ref.~\onlinecite{jpc2}).
Continuous lines refer to the exact results;
dashed lines correspond to the non-polarizable-solute approximation; 
dot-dashed lines have been calculated by taking into account the 
linear polarizability only; dotted lines correspond to the best
single-exponential fit of the exact curve.}
\end{figure}

\end{document}